\documentclass[11pt]{article}

\usepackage[margin=1in]{geometry}
\usepackage{times}
\usepackage{hyperref}
\usepackage[inline]{enumitem}
\usepackage{multirow}
\usepackage{colortbl}
\usepackage{tcolorbox}
\usepackage{graphicx}
\usepackage{amsmath,amssymb}
\usepackage{natbib}
\usepackage{xspace}
\usepackage{pifont}
\usepackage{booktabs}  % for \toprule, \midrule, \bottomrule
\usepackage{array}     % for better column control

\newtcolorbox{grayquote}{
	colback=gray!10,
	colframe=gray!20,
	boxrule=0pt,
	arc=4pt,
	left=2pt,
	right=2pt,
	top=0pt,
	bottom=0pt,
	boxsep=2pt
}

\title{Reassessing Large Language Model Boolean Query Generation for Systematic Reviews}

\author{
	Shuai Wang$^{1}$ \\
	{\small The University of Queensland, Australia} \\
	{\small \texttt{shuai.wang2@uq.edu.au}} \and
	Harrisen Scells$^{2}$ \\
	{\small University of Tübingen, Germany} \\
	{\small \texttt{harrisen.scells@uni-tuebingen.de}} \and
	Bevan Koopman$^{3}$ \\
	{\small CSIRO \& The University of Queensland, Australia} \\
	{\small \texttt{bevan.koopman@csiro.au}} \and
	Guido Zuccon$^{1}$ \\
	{\small The University of Queensland, Australia} \\
	{\small \texttt{g.zuccon@uq.edu.au}}
}

\date{}

\begin{document}
	
	\maketitle
	
	\begin{abstract}
		Systematic reviews are comprehensive literature reviews that address highly focused research questions and represent the highest form of evidence in medicine. A critical step in this process is the development of complex Boolean queries to retrieve relevant literature. Given the difficulty of manually constructing these queries, recent efforts have explored Large Language Models (LLMs) to assist in their formulation. One of the first studies,~\citet{wang2023can}, investigated ChatGPT for this task, followed by~\citet{Moritz2024boolean-reproduce}, which evaluated multiple LLMs in a reproducibility study. However, the latter overlooked several key aspects of the original work, including (i) validation of generated queries, (ii) output formatting constraints, and (iii) selection of examples for chain-of-thought (Guided) prompting. As a result, its findings diverged significantly from the original study. In this work, we systematically reproduce both studies while addressing these overlooked factors. Our results show that query effectiveness varies significantly across models and prompt designs, with guided query formulation benefiting from well-chosen seed studies. Overall, prompt design and model selection are key drivers of successful query formulation. Our findings provide a clearer understanding of LLMs' potential in Boolean query generation and highlight the importance of model- and prompt-specific optimisations. The complex nature of systematic reviews adds to challenges in both developing and reproducing methods but also highlights the importance of reproducibility studies in this domain.
	\end{abstract}
	
	% \keywords{Boolean Query, Systematic Review, Prompt Engineering, LLM} % optional for arXiv

\section{Introduction}
 
Systematic reviews are comprehensive literature reviews focused on a specific research question. They are a cornerstone of evidence-based medicine, playing a critical role in informing decision-making and ensuring the credibility of research findings. To maintain transparency and reproducibility, systematic reviews typically construct a Boolean query. Such queries ensure that the retrieved set of documents remains consistent, enabling others to verify the systematic review process both during peer review and after publication.

Creating an effective Boolean query for systematic review literature search is crucial: it directly controls the quality and cost of the review. A poorly formulated Boolean query could result in too few relevant documents retrieved, thus lowering the quality of the review, or too many non-relevant documents retrieved, increasing the effort (time and cost) of creating the review. Given the importance of an effective Boolean query, systematic review researchers employ specialised librarians called information specialists who can spend weeks formulating complex Boolean queries.

There have been several studies that have employed Generative Large Language Models (LLMs) to assist in the creation of Boolean queries. The first large-scale study was conducted by~\citet{wang2023can}, who investigated whether ChatGPT is capable of creating effective Boolean queries. However, due to the unavailability of APIs at the time (the paper was submitted on January 31, 2023, while the APIs became available on March 1, 2023), their study involved some manual processes to interact with the web chat interface. Key parameters such as the seed, model temperature, and others remain undocumented in this study. A reproducibility study by~\citet{Moritz2024boolean-reproduce} investigated Boolean query generation for systematic review literature search using newer versions of ChatGPT and open-source models such as Mistral~\cite{jiang2023mistral} and Zephyr~\cite{tunstall2023zephyr}. Their study highlighted challenges in replicating the findings of~\citet{wang2023can}, reporting significantly lower recall for generated Boolean queries. Furthermore, one approach in particular from the~\citet{wang2023can} paper—an iterative, chain-of-thought method used in query generation—was criticised as less effective.
 
 \begin{table}[ht]
 	\centering
 	\caption{Summary of our reproducibility study. Key aspects in the original work by~\citet{wang2023can} were overlooked when reproduced by~\citet{Moritz2024boolean-reproduce}. Newer LLMs, in fact, improve the methods originally proposed and show more similar results to manual results, especially in recall. Results from gpt-3.5-1106 (GPT3.5-1). Full results are presented in Tables~\ref{tab:clef-Formulation} and~\ref{tab:seed-Formulation}.}
 		\begin{tabular}{lllllllll}
 			\toprule
 			& & \multirow{2}{*}{Method} & \multicolumn{6}{c}{Prompt} \\ \cmidrule{4-9}
 			& &  & p1 & p2 & p3 & p4 & p5 & Guided \\ 
 			\midrule
 			\multirow{8}{*}{\rotatebox{90}{CLEF TAR collection}} &\multirow{4}{*}{\rotatebox{90}{Precision}} & \multicolumn{7}{c}{Manual = .0209}  \\

 			& & \citet{wang2023can} & .054 & .117 & .084 & .075 & .096 & -- \\  
 			&  & \citet{Moritz2024boolean-reproduce} & \textbf{.345} & \textbf{.319} & \textbf{.320} & \textbf{.284} & \textbf{.290} & -- \\  
 			&  & Ours & .147 & .143 & .125 & .106 & .093 & -- \\  
 			\cmidrule(lr){2-9}
 			&\multirow{4}{*}{\rotatebox{90}{Recall}} & \multicolumn{7}{c}{Manual = .8436} \\
 			&& \citet{wang2023can} & .129 & .131 & .118 & .504 & .334 & --  \\  
 			&  & \citet{Moritz2024boolean-reproduce} & .084 & .067 & .115 & .139 &  .150 & -- \\  
 			&  & Ours & \textbf{.342} & \textbf{.364} & \textbf{.436} & \textbf{.511} & \textbf{.414} &  -- \\  
 			\midrule
 			
 			\multirow{8}{*}{\rotatebox{90}{Seed collection}} &\multirow{4}{*}{\rotatebox{90}{Precision}} & \multicolumn{7}{c}{Manual = .0298}  \\
 			&& \citet{wang2023can} &   \textbf{.051} &  \textbf{.098} &  \textbf{.073} & .028 & .019 &  \textbf{.099} \\  
 			&  & \citet{Moritz2024boolean-reproduce} & .037 & .006 & .036 & .027 &  .037 & .006 \\  
 			&  & Ours & .048 & .057 &.069& \textbf{.077} & \textbf{.093} &.008 \\  
 			\cmidrule(lr){2-9}
 			&\multirow{4}{*}{\rotatebox{90}{Recall}} & \multicolumn{7}{c}{Manual = .7241} \\
 			&& \citet{wang2023can} & .054 & .039 & .052 & .129 & .079 & .517 \\  
 			&  & \citet{Moritz2024boolean-reproduce} & .148 & .025 & .086 & .213 & .244 & .035 \\  
 			&  & Ours & \textbf{.305} & \textbf{.150} & \textbf{.244} & \textbf{.286} & \textbf{.285} & \textbf{.644} \\  
 			\bottomrule
 		\end{tabular}

 	\label{tab:teaser}
 	
 \end{table}
 
We find that the study of~\citet{Moritz2024boolean-reproduce} overlooked several key aspects of the original work by~\citet{wang2023can}, the impact of which is illustrated in Table~\ref{tab:teaser}, namely:
\begin{enumerate*}
	\item \textbf{Query Validation}: The original study checked the validity of generated Boolean queries and regenerated them when a malformed query was generated. The reproducibility study did not account for this step.
	\item \textbf{Query Refinement}: The original study also checked the effectiveness of using ChatGPT for Boolean query refinement, while the reproducibility study solely focused on formulation.
	\item \textbf{Seed Selection}: The original study used the best example for the chain-of-thought approach mentioned above. The reproducibility study used a random example.
	\item \textbf{Prompt and Output Formatting}: The original study used the ChatGPT interface, which does not allow prompt formatting (the use of a system prompt; only a user prompt is defined) and output formatting (only plain-text format instead of JSON format). On the other hand, the reproducibility study put part of the original prompt into the system prompt and constrained the output to JSON~\cite{Moritz2024boolean-reproduce}. We therefore study whether this imposes any effectiveness difference in the Boolean queries generated, which was pointed out in previous research: restricting output type could reduce the effectiveness of LLM generation~\cite{tam2024let}.
\end{enumerate*}
 
This paper addresses these key differences to provide a more systematic investigation into using LLMs for Boolean query generation for systematic review literature search. Specifically, we seek to (1) revisit the approaches from~\citet{wang2023can} and rerun all the models from the reproducibility study of~\citet{Moritz2024boolean-reproduce}, incorporating query validation to ensure that effectiveness is not impacted by queries that are malformed and thus retrieve zero documents; (2) compare unrestricted output instructions, JSON-restricted output instructions, and the use of a system prompt to investigate whether the changes in the reproducibility impact effectiveness; (3) further investigate the impact of seed selection strategies, including best example and combined seed studies; and
(4) conduct a case study showcasing actual examples of Boolean queries formulated by different prompts or models, and analyze their impact on retrieval effectiveness.
 
	\section{Related Work}

\subsection{Automatic Boolean Query Formulation}

\enlargethispage{1\baselineskip}
Current automated Boolean query formulation methods fall into two broad categories. The first relies on statistical techniques inspired by the conceptual and objective methods, employing heuristics to combine and join terms automatically to formulate Boolean queries~\citep{scells2020conceptual,scells2020objective}. The second category leverages the generative capabilities of Large Language Models (LLMs) to produce Boolean queries directly from input information, such as review topics and seed studies~\citep{wang2023can, Moritz2024boolean-reproduce}.

\vspace{-2pt}
\subsubsection{Statistical Approaches}
Research in this area has consistently found that fully automated procedures do not match human performance; however, manual refinement of automatically generated queries can significantly improve retrieval effectiveness. Query formulation is the process of deriving a Boolean query from a research question based on established guidelines. Two well-recognized procedures have emerged for developing queries in systematic reviews. The first, known as the \textit{conceptual} method~\cite{suhail2013methods}, involves initially identifying high-level concepts from pilot searches or pre-identified relevant studies. These concepts then guide the discovery of synonyms and related keywords, with the query iteratively refined by an information specialist. The second, termed the \textit{objective} method~\cite{simon2010identifying,hausner2015development}, begins with a small set of potentially relevant studies that seed the process. This approach employs statistical procedures to extract terms from these studies, providing an initial query structure that is further refined manually by a specialist. While these methods differ in their starting points, both ultimately rely on expert intervention for query optimization. 

\vspace{-2pt}
\subsubsection{Generative Approaches}
Research in using LLMs for query formulation diverges in their results. For example, the first study that used the ChatGPT interface for formulation suggested that LLMs could generate Boolean queries obtaining a recall of approximately 0.5 on average in automated formulation prompts for CLEF TAR topics, but the subsequent reproducibility study on the paper, under various API-based and open-sourced LLMs, found that the best recall achieved for the same dataset was only 0.15 on average~\cite{wang2023can, Moritz2024boolean-reproduce}. This substantial discrepancy raises concerns about the reliability of LLM-generated queries, particularly in maintaining the high recall required for systematic review search strategies.

Beyond Boolean query formulation, generative models have also been explored for translating complex Boolean queries into natural language queries. These natural language representations can then be combined with effective ranking models to enhance document screening in systematic review creation~\citep{wang2023generating}. This approach provides an alternative way to leverage LLMs in systematic review workflows by improving retrieval effectiveness without relying solely on Boolean query formulation.

\subsection{Automatic Boolean Query Refinement}

Query refinement involves improving an initial query automatically. Statistical and feature-based approaches have been explored to iteratively refine queries based on retrieval performance~\citep{scells2018generating, scells2019refine}. Other techniques have involved modifying the function of the Boolean operators themselves~\cite{scells2023smooth}. Methods have also been developed to refine specific groups of terms within Boolean queries, such as MeSH term suggestions~\citep{wang2021mesh, wang2022automated, wang2022meshsuggester}. These approaches aim to balance retrieval effectiveness by reducing irrelevant results while maintaining comprehensive coverage of relevant studies.

More recent research has investigated the use of generative models for query refinement. \citet{wang2023can} investigated the use of ChatGPT for refining Boolean queries, evaluating its ability to enhance queries that were either manually created by search experts or generated using statistical methods. The study found that automatic Boolean query refinement can improve query precision by eliminating noisy or less relevant terms; however, this often comes at the cost of reduced recall.

%\cite{scells2018generating,scells2019automatic} using statistical, feature-based method to automatically refine query

%\cite{scells2018qvpp} uses query performance prediction to identify the most effective natural language query for systematic review literature search

%\cite{wang2021mesh} refining the mesh terms in queries to make them more precise

%\cite{wang2023can,Moritz2024boolean-reproduce} both investigate query refinement using generative models

%\vfill\eject

	\section{Experimental Setup}

\subsection{Datasets}
Our experiments closely follow the two studies of~\citet{wang2023can} and~\citet{Moritz2024boolean-reproduce}. We use two primary collections: (1)the CLEF Technology-Assisted Reviews (CLEF TAR) collections\cite{kanoulas2017clef, kanoulas2018clef}, and (2)the Seed Collection\cite{wang2022seed}. For CLEF TAR, we combined topics from CLEF TAR 2017 and 2018. After removing duplicate topics (\texttt{CD010771}, \texttt{CD011145}, \texttt{CD010772}, \texttt{CD010775}, \texttt{CD010783}, \texttt{CD010896}, \texttt{CD007431}, and \texttt{CD010860}), 72 unique topics remain. For the Seed Collection, we use two versions:
\begin{enumerate*}
	\item \textbf{Original version:} The collection as used by~\citet{wang2023can}, consisting of 40 topics. Each topic represents a distinct search task, potentially covering different aspects of the same topic, such as varying focus points or seed studies.
	\item \textbf{Deduplicated version:} A version of the collection used by~\citet{Moritz2024boolean-reproduce}, which removes duplicate review searches. Each topic corresponds to a distinct systematic review, resulting in a total of 35 unique systematic reviews. Specifically, topics 51, 52, and 53 were merged, as were topics 43 and 96, 7 and 67, and 8 and 112. The included studies and seed studies were combined accordingly, while the topic titles and search dates remained unchanged.
\end{enumerate*}

\subsection{Prompts}
We use the same prompts employed in the original and reproduction studies:
\begin{enumerate*}
	\item \textbf{Query formulation prompts (p1-p5):} p1 to p3 are zero-shot prompts that directly ask the LLM to generate a Boolean query by providing the review topic, while p4 and p5 are one-shot prompts and are provided a high-quality example query from the collection.
	\item \textbf{Query refinement prompts (p6-p7):} These prompts are used to refine existing Boolean queries. p6 is a zero-shot prompt, while p7 is a one-shot prompt and is provided a high-quality example query from the collection.
	\item \textbf{Guided query formulation prompt:} This chain-of-thought prompt generates Boolean queries using seed studies.
\end{enumerate*}
For example-based prompts, we followed the original study and used the high-quality example query \texttt{CD010438}.

\subsection{Models}
We include and extend the set of models used in the reproducibility study, incorporating the following:
\begin{enumerate*}
	\item \textbf{OpenAI:} GPT-3.5-turbo-1106 (GPT3.5-1), GPT-3.5-turbo-0125 (GPT3.5-0), GPT-4-1106-preview (GPT4), GPT-4o-mini (GPT4o-m)
	\item \textbf{Mistral:} Mistral-7B-v0.2 (Mistral-S), Mixtral-8x7B-v0.1 (Mistral-L)
	\item \textbf{Local:} Mistral-7B-v0.2 (Mistral), Zephyr-7B-beta (Zephyr), Llama3.1-8B-Instruct (Llama3.1)
	\item \textbf{Reasoning:} OpenAI-o1 (o1)
\end{enumerate*}

In total, we used ten different LLMs. This includes four additional models beyond those in the reproducibility study. For all models, we set the temperature to 1 to enable variation in Boolean query formulation upon regeneration. Additionally, we set the random seed to 42 to ensure outputs remained consistent.

\subsection{Query Validation}
The original study involved manual removal of incorrectly generated queries, whereas the reproducibility study did not validate queries. To ensure consistency with the original study, we improved this process by implementing an automated query validation pipeline consisting of three steps:
\begin{enumerate*}
	\item \textbf{LLM-based extractor:} Since LLM-generated outputs may contain explanations or reasoning alongside the Boolean query, we employed an LLM-based extractor to isolate the Boolean query component. For consistency, we always used GPT-3.5-turbo-0125 with a temperature of 0.
	\item \textbf{Rule-based checker:} Extracted Boolean queries were passed through a rule-based checker to ensure compliance with Boolean syntax. We enforce the following rules: (1) Every bracket must be closed. (2) Only three Boolean operators are allowed: “NOT”, “AND”, “OR”. (3) Each operation must use a single operator at a time; consecutive operators (e.g., “AND OR”) are invalid. (4) Each operator must be preceded by a term or an opening bracket.
	\item \textbf{Boolean query validator:} We used PubMed’s Entrez API~\cite{sayers2010general} to verify whether the automatically generated Boolean queries retrieved a reasonable number of documents. A query was considered valid if it retrieved between one and one million documents~\footnote{The upper bound of one million documents is primarily determined by the computation budget for API calls; in practice, this number should be set according to actual needs.}. All Boolean queries were executed with the original search date to ensure consistency with the time the topics were created.
\end{enumerate*}
To prevent excessive query regeneration, we set a maximum number of 20 attempts per topic. If a valid query was not produced within 20 attempts, we used the last generated query regardless of its validity.

\subsection{Evaluation}
To evaluate the generated Boolean queries, we executed them via PubMed’s Entrez API, retrieving PubMed IDs~\cite{canese2013pubmed}. The retrieved PubMed IDs were compared against abstract-level relevance judgments in the collection using set-based evaluation measures, including precision, recall, and F-measure.

	\section{Reproducing LLM-based Boolean Query Formulation and Refinement}

\begin{table*}[ht]
	\centering
	\caption{Results on 71 CLEF topics for all query-formulation prompts.
		For each evaluation metric, \textbf{bolded} values indicate the highest value among all prompts for a given large language model, and \textcolor{gray}{coloured} values indicate the highest effectiveness among all model variations within each prompt.
		A paired t-test with Bonferroni correction ($p < .05$) is performed:
		${a}$ indicates statistical significance relative to the \textbf{bolded} value,
		${b}$ relative to the \textcolor{gray}{coloured} value,
		and ${c}$ relative to the manual baseline.}
	\resizebox{\textwidth}{!}{%
		\begin{tabular}{llllllllllllll}
			\toprule
			& Prompt & GPT3.5-1 & GPT3.5-0 & GPT4 & GPT4o-m & Mistral-S & Mistral-L & Mistral & Zephyr & Llama3.1 & o1 & Wang & Staudinger \\ \midrule
			\multirow{6}{*}{\rotatebox[origin=c]{90}{Precision}} & \multicolumn{13}{c}{Manual = .0209} \\
			& p1 & \textbf{\cellcolor{lightgray}{.1470}}$^{c}$ & .1451$^{c}$ & .0854$^{a}$ & .1393$^{c}$ & .0139$^{a,b}$ & .0384$^{a,b}$ & .0415$^{b}$ & .0792 & .1069$^{c}$ & \textbf{.1361}$^{c}$ & .0543$^{b}$ & .345 \\
			& p2 & .1430$^{c}$ & \textbf{.1501}$^{c}$ & \textbf{\cellcolor{lightgray}{.1707}}$^{c}$ & .1294$^{c}$ & \textbf{.1008} & \textbf{.1329}$^{c}$ & \textbf{.1233}$^{c}$ & .1088$^{c}$ & .1097$^{c}$ & .1324$^{c}$ & \textbf{.1166}$^{c}$ & .319\\
			& p3 & .1249$^{c}$ & .1051$^{c}$ & .1348$^{c}$ & \textbf{\cellcolor{lightgray}{.1409}}$^{c}$ & .0617$^{b}$ & .1019$^{c}$ & .0732$^{c}$ & .1057 & \textbf{.1153}$^{c}$ & .1078$^{c}$ & .0844$^{c}$ & .320\\
			& p4 & .1064$^{c}$ & .1006$^{c}$ & \cellcolor{lightgray}{.1333}$^{c}$ & .0662$^{b}$ & .0844 & .0965$^{c}$ & .0624$^{b}$ & .1135$^{c}$ & .0685 & .1060$^{c}$ & .0752 & .284\\
			& p5 & .0934$^{c}$ & .1190$^{c}$ & \cellcolor{lightgray}{.1372}$^{c}$ & .0971$^{c}$ & .0803$^{c}$ & .1109$^{c}$ & .0904$^{c}$ & \textbf{.1319}$^{c}$ & .0801 & .1189$^{c}$ & .0958$^{c}$ & .290 \\
			\midrule
			\multirow{6}{*}{\rotatebox[origin=c]{90}{Recall}} & \multicolumn{13}{c}{Manual = .8436} \\
			& p1 & .3419$^{a,c}$ & .2529$^{a,b,c}$ & .1183$^{a,b,c}$ & .3235$^{a,c}$ & .0406$^{a,b,c}$ & .1533$^{a,b,c}$ & .0754$^{a,b,c}$ & .0410$^{a,b,c}$ & .1899$^{a,b,c}$ & \cellcolor{lightgray}{.4341}$^{a,c}$ & .1293$^{a,b,c}$ & .084\\
			& p2 & .3643$^{c}$ & .2808$^{a,b,c}$ & .2595$^{b,c}$ & .2878$^{a,b,c}$ & .1829$^{a,b,c}$ & .2484$^{a,b,c}$ & .0895$^{a,b,c}$ & .1554$^{b,c}$ & .3753$^{a,c}$ & \cellcolor{lightgray}{.5252}$^{a,c}$ & .1310$^{a,b,c}$  &.067 \\
			& p3 & .4357$^{b,c}$ & .4120$^{b,c}$ & .3112$^{b,c}$ & .3343$^{a,b,c}$ & .1137$^{a,b,c}$ & \textbf{.4304}$^{b,c}$ & .2049$^{a,b,c}$ & .1691$^{b,c}$ & .3741$^{a,b,c}$ & \textbf{\cellcolor{lightgray}{.6545}}$^{c}$ & .1175$^{a,b,c}$ & .115 \\
			& p4 & \textbf{.5108}$^{c}$ & \textbf{.4862}$^{c}$ & .3149$^{b,c}$ & \textbf{.5509}$^{c}$ & \textbf{.4831}$^{c}$ & .3615$^{b,c}$ & .4345$^{c}$ & .2655$^{b,c}$ & \textbf{\cellcolor{lightgray}{.5811}}$^{c}$ & .5438$^{a,c}$ & \textbf{.5035}$^{c}$ & .139\\
			& p5 & .4138$^{c}$ & .4654$^{c}$ & \textbf{.3200}$^{b,c}$ & .4576$^{c}$ & .3174$^{b,c}$ & .3191$^{b,c}$ & \textbf{.4833}$^{c}$ & \textbf{.2904}$^{b,c}$ & .4740$^{c}$ & \cellcolor{lightgray}{.5377}$^{a,c}$ & .3335$^{b,c}$ & .150 \\
			\midrule
			\multirow{6}{*}{\rotatebox[origin=c]{90}{F1}} & \multicolumn{13}{c}{Manual = .0298}  \\
			& p1 & .1083$^{c}$ & .0970$^{c}$ & .0551$^{a,b}$ & .1020$^{c}$ & .0118$^{a,b}$ & .0381$^{a,b}$ & .0226$^{a,b}$ & .0252$^{a,b}$ & .0665$^{b}$ & \cellcolor{lightgray}{.1335}$^{c}$ & .0500$^{b}$ & .097 \\
			& p2 & \textbf{.1083}$^{c}$ & .0981$^{c}$ & .1016$^{c}$ & .1073$^{c}$ & .0536$^{b}$ & .0839$^{b,c}$ & .0414$^{a,b}$ & .0593$^{b}$ & .0875$^{c}$ & \cellcolor{lightgray}{.1427}$^{c}$ & .0654$^{b,c}$ & .084 \\
			& p3 & .1053$^{c}$ & .0922$^{c}$ & .1043$^{c}$ & \textbf{.1108}$^{c}$ & .0363$^{b}$ & .0845$^{b,c}$ & .0446$^{a,b}$ & .0401$^{a,b}$ & \textbf{.1045}$^{c}$ & \cellcolor{lightgray}{.1381}$^{c}$ & .0443$^{b}$ & .101\\
			& p4 & .0996$^{c}$ & .0818$^{b,c}$ & \textbf{.1135}$^{c}$ & .0525$^{a,b}$ & .0617$^{b}$ & .0731 & .0586$^{b}$ & .0734$^{b,c}$ & .0673$^{b}$ & \cellcolor{lightgray}{.1316}$^{c}$ & .0642$^{b}$ & .123\\
			& p5 & .0798$^{b,c}$ & \textbf{.1017}$^{c}$ & .1092$^{c}$ & .0809$^{b,c}$ & \textbf{.0666}$^{b}$ & \textbf{.0937}$^{b,c}$ & \textbf{.0860}$^{b,c}$ & \textbf{.0928}$^{c}$ & .0617$^{b}$ & \textbf{\cellcolor{lightgray}{.1428}}$^{c}$ & \textbf{.0717}$^{b,c}$ & .134\\
			\midrule
			\multirow{6}{*}{\rotatebox[origin=c]{90}{F3}} & \multicolumn{13}{c}{Manual = .0497} \\
			& p1 & .1311$^{c}$ & .1114$^{c}$ & .0608$^{a,b}$ & .1269$^{c}$ & .0128$^{a,b,c}$ & .0495$^{a,b}$ & .0275$^{a,b}$ & .0265$^{a,b}$ & .0746$^{b}$ & \cellcolor{lightgray}{.1686}$^{c}$ & .0590$^{b}$&  -- \\
			& p2 & .1338$^{c}$ & .1147$^{b,c}$ & .1206$^{c}$ & .1252$^{c}$ & .0616$^{b}$ & .0966$^{b}$ & .0428$^{a,b}$ & .0674$^{b}$ & .1110$^{c}$ & \cellcolor{lightgray}{.1859}$^{c}$ & .0696$^{b}$&  --\\
			& p3 & \textbf{.1363}$^{c}$ & .1181$^{b,c}$ & .1275$^{b,c}$ & \textbf{.1341}$^{b,c}$ & .0437$^{b}$ & .1025$^{b,c}$ & .0539$^{a,b}$ & .0534$^{a,b}$ & \textbf{.1301}$^{b,c}$ & \textbf{\cellcolor{lightgray}{.1966}}$^{c}$ & .0497$^{b}$&  -- \\
			& p4 & .1335$^{c}$ & .1049$^{b,c}$ & \textbf{.1392}$^{c}$ & .0631$^{a,b}$ & \textbf{.0839}$^{b}$ & .0863$^{b}$ & .0822$^{b}$ & .0877$^{b}$ & .0913$^{b}$ & \cellcolor{lightgray}{.1777}$^{c}$ & \textbf{.0847}$^{b}$&  -- \\
			& p5 & .1064$^{b,c}$ & \textbf{.1250}$^{b,c}$ & .1364$^{b,c}$ & .1085$^{b,c}$ & .0800$^{b}$ & \textbf{.1141}$^{b,c}$ & \textbf{.1139}$^{b,c}$ & \textbf{.1077}$^{b,c}$ & .0829$^{b}$ & \cellcolor{lightgray}{.1959}$^{c}$ & .0844$^{b}$&  --\\
			\bottomrule
		\end{tabular}
	}
	\label{tab:clef-Formulation}
\end{table*}

\begin{table*}[ht]
	\centering
	\caption{Results on 40 Seed collection topics for all query-formulation prompts.
		For each evaluation metric, \textbf{bolded} values indicate the highest value among all prompts for a given large language model, and \textcolor{gray}{coloured} values indicate the highest effectiveness among all model variations within each prompt.
		A paired t-test with Bonferroni correction ($p < .05$) is performed:
		${a}$ indicates statistical significance relative to the \textbf{bolded} value,
		${b}$ relative to the \textcolor{gray}{coloured} value,
		and ${c}$ relative to the manual baseline.
		The o1 model for the Guided prompt was not evaluated due to high expenses.}
	\resizebox{0.97\textwidth}{!}{%
		\begin{tabular}{llllllllllllll}
			\toprule
			& Prompt & GPT3.5-1 & GPT3.5-0 & GPT4 & GPT4o-m & Mistral-S & Mistral-L & Mistral & Zephyr & Llama3.1 & o1 & Wang & Staudinger \\ \midrule
			\multirow{8}{*}{\rotatebox[origin=c]{90}{Precision}} & \multicolumn{13}{c}{Manual = .0341} \\
			& p1 & .0475 & .0309 & .0262$^{b}$ & \textbf{.0491} & .0090 & .0079 & .0155 & .0431 & \textbf{.0605} & \textbf{\cellcolor{lightgray}{.0827}} & .0514 & .037 \\
			& p2 & .0573 & .0251 & \textbf{.0344} & .0151 & \textbf{.0962} & \textbf{.0470} & .0355 & .0942 & .0599 & .0469 & \cellcolor{lightgray}{.0983} & .006 \\
			& p3 & .0687 & \textbf{\cellcolor{lightgray}{.0757}} & .0313 & .0370 & .0680 & .0451 & \textbf{.0706} & .0255 & .0349 & .0523 & .0730 & .036 \\
			& p4 & \cellcolor{lightgray}{.0768} & .0602 & .0126 & .0337 & .0142 & .0378 & .0178 & .0282 & .0132 & .0384 & .0284 & .027\\
			& p5 & \textbf{.0932} & .0229 & .0011$^{c}$ & .0145 & .0611 & .0466 & .0167 & \textbf{\cellcolor{lightgray}{.1104}} & .0202 & .0618 & .0189 & .037\\
			\cmidrule{2-14}
			& Guided-best & .0078 & .0061 & .0209 & .0210 & .0497 & .0292 & .0381 & .0755 & .0158 & -- & \textbf{\cellcolor{lightgray}{.0993}} & .099\\
			& Guided-combined & .0002$^{c}$ & .0010$^{c}$ & .0022$^{c}$ & .0071 & .0038$^{c}$ & \cellcolor{lightgray}{.0165} & .0015$^{c}$ & .0087 & .0022$^{c}$ & -- & -- & -- \\
			\midrule
			\multirow{8}{*}{\rotatebox[origin=c]{90}{Recall}} & \multicolumn{13}{c}{Manual = .7241} \\
			& p1 & .3045$^{a,c}$ & .1318$^{a,b,c}$ & .0600$^{a,b,c}$ & .1930$^{a,c}$ & .0898$^{a,b,c}$ & .1142$^{a,b,c}$ & .0244$^{a,b,c}$ & .0321$^{a,b,c}$ & .2446$^{a,c}$ & \cellcolor{lightgray}{.3767}$^{a,c}$ & .0542$^{a,b,c}$ & .148 \\
			& p2 & .1495$^{a,c}$ & .0858$^{a,b,c}$ & .0580$^{a,b,c}$ & .1625$^{a,c}$ & .0644$^{a,b,c}$ & .1614$^{a,c}$ & .0858$^{a,b,c}$ & .0876$^{a,b,c}$ & .2177$^{a,c}$ & \cellcolor{lightgray}{.3245}$^{a,c}$ & .0394$^{a,b,c}$ & .025\\
			& p3 & .2437$^{a,b,c}$ & .1986$^{a,b,c}$ & .0321$^{a,b,c}$ & .0982$^{a,b,c}$ & .1154$^{a,b,c}$ & .2774$^{a,b,c}$ & .0991$^{a,b,c}$ & .0321$^{a,b,c}$ & .2329$^{a,b,c}$ & \textbf{\cellcolor{lightgray}{.5786}} & .0519$^{a,b,c}$ & .086 \\
			& p4 & \cellcolor{lightgray}{.2860}$^{a,c}$ & .2078$^{a,c}$ & .0056$^{a,b,c}$ & .2077$^{a,c}$ & .1580$^{a,c}$ & .1180$^{a,c}$ & .1536$^{a,c}$ & .1682$^{a,c}$ & .2251$^{a,c}$ & .2592$^{a,c}$ & .1290$^{a,c}$ & .213\\
			& p5 & .2848$^{a,c}$ & .3285$^{a,c}$ & .0231$^{a,b,c}$ & .2746$^{a,c}$ & .2385$^{a,c}$ & .1251$^{a,b,c}$ & .2986$^{a,c}$ & .1928$^{a,c}$ & .3113$^{a,c}$ & \cellcolor{lightgray}{.4182}$^{a,c}$ & .0785$^{a,b,c}$ & .244\\
			\cmidrule{2-14}
			& Guided-best & \cellcolor{lightgray}{.6441}$^{a}$ & .6328$^{a}$ & .6188$^{a}$ & .3820$^{a,b,c}$ & .6091$^{a}$ & .5042$^{a,b}$ & .5241$^{a}$ & .4434$^{a,b,c}$ & .6070$^{a}$ & -- & \textbf{.5171} & .035\\
			& Guided-combined & \textbf{\cellcolor{lightgray}{.7361}} & \textbf{.7315} & \textbf{.6978} & \textbf{.5045}$^{b}$ & \textbf{.6649} & \textbf{.5908} & \textbf{.5772} & \textbf{.5134}$^{b}$ & \textbf{.6801} & -- & -- & --\\
			\midrule
			\multirow{8}{*}{\rotatebox[origin=c]{90}{F1}} & \multicolumn{13}{c}{Manual = .0605} \\
			& p1 & .0410 & .0221$^{b}$ & .0071$^{b,c}$ & \textbf{.0441} & .0140$^{b,c}$ & .0118$^{b,c}$ & .0131$^{b}$ & .0167 & \textbf{.0627} & \textbf{\cellcolor{lightgray}{.0777}} & .0281$^{b}$ & .038 \\
			& p2 & .0322 & .0227 & .0111$^{c}$ & .0205 & \textbf{.0424} & .0323 & .0220 & .0246 & .0436 & \cellcolor{lightgray}{.0533} & .0310 & .003\\
			& p3 & \cellcolor{lightgray}{.0597} & \textbf{.0525} & .0128 & .0198 & .0385 & \textbf{.0357} & \textbf{.0439} & .0224 & .0421 & .0537 & .0329 & .019 \\
			& p4 & \cellcolor{lightgray}{.0471} & .0383 & .0047$^{c}$ & .0169 & .0123$^{c}$ & .0143 & .0190 & .0252 & .0159$^{c}$ & .0262 & .0274 & .023 \\
			& p5 & \textbf{.0612} & .0355 & .0018$^{b,c}$ & .0184 & .0347 & .0299 & .0253$^{c}$ & \textbf{\cellcolor{lightgray}{.0698}} & .0267 & .0561 & .0193 & .032 \\
			\cmidrule{2-14}
			& Guided-best & .0139 & .0110$^{c}$ & \textbf{.0335} & .0304 & .0334 & .0232 & .0300 & .0441 & .0216 & -- & \textbf{\cellcolor{lightgray}{.0492}} & .002\\
			& Guided-combined & .0003$^{a,c}$ & .0020$^{a,c}$ & .0044$^{c}$ & \cellcolor{lightgray}{.0100} & .0046$^{c}$ & .0040$^{c}$ & .0008$^{c}$ & .0083$^{a,c}$ & .0035$^{c}$ & -- & -- & --\\
			\midrule
			\multirow{8}{*}{\rotatebox[origin=c]{90}{F3}} & \multicolumn{13}{c}{Manual = .1024} \\
			& p1 & .0571 & .0266$^{b,c}$ & .0069$^{b,c}$ & \textbf{.0534} & .0198$^{b,c}$ & .0175$^{b,c}$ & .0143$^{b,c}$ & .0175$^{b,c}$ & \textbf{.0769} & \textbf{\cellcolor{lightgray}{.1019}} & .0306$^{b}$& -- \\
			& p2 & .0388 & .0273$^{c}$ & .0108$^{c}$ & .0304$^{c}$ & .0400 & .0376 & .0211$^{c}$ & .0241$^{c}$ & .0544 & \cellcolor{lightgray}{.0755} & .0278$^{c}$ & --\\
			& p3 & \textbf{.0820} & \textbf{.0652} & .0141$^{b,c}$ & .0242$^{c}$ & .0416 & \textbf{.0512} & \textbf{.0418} & .0240 & .0555 & \cellcolor{lightgray}{.0852} & .0329& -- \\
			& p4 & \cellcolor{lightgray}{.0566} & .0469 & .0038$^{b,c}$ & .0228$^{c}$ & .0151$^{c}$ & .0133$^{c}$ & .0243$^{c}$ & .0338$^{c}$ & .0225$^{c}$ & .0366$^{c}$ & .0374& -- \\
			& p5 & \cellcolor{lightgray}{.0791} & .0526 & .0027$^{b,c}$ & .0302$^{c}$ & \textbf{.0440} & .0375 & .0373$^{c}$ & \textbf{.0783} & .0352 & .0780 & .0271$^{c}$& -- \\
			\cmidrule{2-14}
			& Guided-best & .0232$^{c}$ & .0187$^{c}$ & \textbf{.0509} & .0420 & .0424 & .0363 & .0360 & \cellcolor{lightgray}{.0581} & .0314$^{c}$ & -- & \textbf{.0565} & --\\
			& Guided-combined & .0006$^{a,c}$ & .0036$^{a,c}$ & .0084$^{c}$ & .0132$^{c}$ & .0083$^{c}$ & .0064$^{c}$ & .0011$^{c}$ & \cellcolor{lightgray}{.0134}$^{a,c}$ & .0057$^{c}$ & -- & -- & --\\
			\bottomrule
		\end{tabular} 
	}
	\label{tab:seed-Formulation}
\end{table*}

The original paper categorises prompts into three types—formulation, guided formulation, and refinement. In this section, we present the results of our reproduction using these three prompts.

\subsection{Formulation Prompts}

The query formulation results on the CLEF TAR and Seed collections are presented in Tables~\ref{tab:clef-Formulation} and~\ref{tab:seed-Formulation}, respectively. Comparing our reproduced results to the original study by~\citet{wang2023can}, which used an early version of ChatGPT, we observe that most LLM variants achieve higher effectiveness in both recall and precision.

The tables also include results for GPT3.5-1 from the reproducibility study.\footnote{Due to space constraints, we provide a detailed comparison in the GitHub repository: \url{https://github.com/ielab/Boolean_Generation_Reproduce}} We note a larger difference between GPT3.5-1’s performance in that study and in our reproduced results. Specifically, the recall reported by~\citet{Moritz2024boolean-reproduce} is much lower across nearly all metrics for most model variants and prompts. This discrepancy highlights one of the issues with the reproducibility study: it does not verify the validity of generated Boolean queries, leading to many invalid queries that retrieve no results.

The most effective model or prompt in our reproduced results depends on the evaluation metric and the dataset used. When comparing different prompts, zero-shot formulation prompts (p1–p3) generally obtain higher precision but lower recall compared to one-shot formulation prompts (p4–p5). This results in mixed F-metric values. Compared to the original results, out of ten models evaluated, one-shot prompts achieve higher effectiveness on seven models for F1 and six models for F3 on the CLEF TAR collection; they are also more effective than the others in two models for both F1 and F3 on the Seed collection.

When comparing models using the same prompt, o1 (the most recent reasoning model) consistently obtains higher recall than other variants, except when p4 is used. In that instance, o1 shows slightly lower recall than Llama3.1 on the CLEF TAR collection and GPT3.5-1 on the Seed collection, although the differences are not statistically significant. On the CLEF TAR collection, o1 also outperforms other models in all prompt variations for F-based metrics. Among the remaining models, Llama3.1 and GPT3.5-1 obtain relatively higher recall under several prompt settings; both have the top recall in five instances (combining CLEF TAR and Seed collections across the five prompt variations).

Compared to the manually formulated Boolean queries across both collections, our findings confirm the claim by~\citet{wang2023can} that automatic methods obtain higher precision but significantly lower recall (with the only exception being o1 on the Seed collection using p3). Nevertheless, recall has notably improved since the study by~\citet{wang2023can}, particularly for the most recent o1 model. Average recall is now only approximately 15\% lower than that of the manual queries on both collections, whereas the original study reported a 35\% gap on CLEF and a 60\% gap on the Seed collection.

\subsection{Guided Formulation Prompts}

Table~\ref{tab:seed-Formulation} also contains the results of guided query formulation. Our results, shown as Guided-best, support the original claim that with a well-chosen seed study, queries generated from the guided prompt are more effective than those generated from the unguided prompt. More specifically, our findings highlight that selecting the best seed study for guided prompts (Guided-best) can significantly increase recall compared to all one-step query formulation prompts from the previous subsection across all the models evaluated.

The results of our guided query formulation align with~\citet{wang2023can}. Recall was improved for six out of nine models. We did not test the o1 model due to high costs. When compared to the original queries, most models do not exhibit a statistically significant difference in effectiveness. The exceptions are Mistral-S for recall, GPT3.5-0 for F1, and GPT3.5-0, GPT3.5-1, and Llama3.1 for F3.

We also combined all queries generated from different seed studies with OR, shown as Guided-combined. These results demonstrate that combining queries in this way can achieve recall levels comparable to those of the original Boolean query, with no statistically significant difference for all tested models. However, this approach comes with a trade-off of lower precision, which is usually statistically significantly different from the original Boolean query for all models except Zephyr, Mistral-L, and GPT4o-mini. Notably, applying a combined seed strategy is a pre-hoc approach, meaning there is no need to choose a single best seed study as in Guided-best.

\subsection{Refinement Prompts}

\begin{table*}[ht]
	\centering
		\caption{Results on 71 CLEF TAR topics for all query-refinement prompts.
			For each evaluation metric, \textbf{bolded} values indicate the highest value among all prompts for a given large language model, and \textcolor{gray}{coloured} values indicate the highest effectiveness among all model variations within each prompt.
			A paired t-test with Bonferroni correction ($p < .05$) is performed:
			${a}$ indicates statistical significance relative to the \textbf{bolded} value (highest prompt for each model),
			${b}$ relative to the \textcolor{gray}{coloured} value,
			and ${c}$ relative to the refinement base, i.e., p6-Manual and p6, p7-conceptual and p7, and p7-objective and objective.}
		\label{tab:clef-Refinement}
	\resizebox{\textwidth}{!}{%
		\begin{tabular}{llllllllllll}
			\toprule
			& Prompt & GPT3.5-1 & GPT3.5-0 & GPT4 & GPT4o-m & Mistral-S & Mistral-L & Mistral & Zephyr & Llama3.1 & Wang \\
			\midrule
			\multirow{4}{*}{\rotatebox[origin=c]{90}{Precision}} & \multicolumn{11}{c}{Manual=.0209, conceptual=.0015, objective=.0002} \\
			& p6-Manual & .0393$^{a,b,c}$ & .0492$^{b,c}$ & .1211$^{c}$ & \textbf{.0812}$^{b,c}$ & \textbf{.0873}$^{c}$ & .0839$^{a,c}$ & \textbf{.0865}$^{c}$ & \textbf{\cellcolor{lightgray}.1234}$^{c}$ & \textbf{.1186}$^{c}$ & \textbf{.0795}$^{c}$ \\
			& p7-conceptual & \textbf{.0982}$^{b,c}$ & \textbf{.0742}$^{b,c}$ & \textbf{\cellcolor{lightgray}.1581}$^{c}$ & .0547$^{b,c}$ & .0575$^{b,c}$ & \textbf{.1411}$^{c}$ & .0596$^{b,c}$ & .0583$^{a,b,c}$ & .0566$^{a,b,c}$ & .0022$^{a,b}$ \\
			& p7-objective & .0460$^{a,b,c}$ & .0242$^{a,b,c}$ & \cellcolor{lightgray}.1100$^{a,c}$ & .0041$^{a,b,c}$ & .0317$^{a,b,c}$ & .0906$^{c}$ & .0237$^{a,b,c}$ & .0416$^{a,b,c}$ & .0135$^{a,b,c}$ & .0460$^{b,c}$ \\
			\midrule
			\multirow{4}{*}{\rotatebox[origin=c]{90}{Recall}} & \multicolumn{11}{c}{Manual=.8436, conceptual=.6996, objective=.9128} \\
			& p6-Manual & \textbf{.6653}$^{c}$ & \cellcolor{lightgray}.7057$^{c}$ & .3934$^{b,c}$ & .4828$^{a,b,c}$ & .3944$^{a,b,c}$ & .4346$^{b,c}$ & .4340$^{a,b,c}$ & .2450$^{a,b,c}$ & .3833$^{a,b,c}$ & .5060$^{a,b,c}$ \\
			& p7-conceptual & .4630$^{a,b,c}$ & .4744$^{a,c}$ & .2834$^{a,b,c}$ & .5092$^{a,c}$ & .4735$^{a,c}$ & .3120$^{a,b,c}$ & .5548$^{c}$ & .3004$^{a,b,c}$ & \cellcolor{lightgray}.5827$^{a,c}$ & .2699$^{a,b,c}$ \\
			& p7-objective & .5763$^{b,c}$ & \textbf{.7175}$^{c}$ & \textbf{.4339}$^{b,c}$ & \textbf{.8078}$^{c}$ & \textbf{.7386}$^{c}$ & \textbf{.5011}$^{b,c}$ & \textbf{.6188}$^{b,c}$ & \textbf{.5517}$^{b,c}$ & \textbf{.7951}$^{c}$ & \textbf{\cellcolor{lightgray}.8115}$^{c}$ \\
			\midrule
			\multirow{4}{*}{\rotatebox[origin=c]{90}{F1}} & \multicolumn{11}{c}{Manual=.0298, conceptual=.0027, objective=.0005} \\
			& p6-Manual & .0536$^{a,b,c}$ & \textbf{.0648}$^{b,c}$ & \cellcolor{lightgray}.1149$^{c}$ & \textbf{.0882}$^{c}$ & \textbf{.0715}$^{b,c}$ & .0786$^{b,c}$ & \textbf{.0659}$^{b,c}$ & \textbf{.0810}$^{b,c}$ & \textbf{.0840}$^{b,c}$ & \textbf{.0597}$^{b,c}$ \\
			& p7-conceptual & \textbf{.0882}$^{b,c}$ & .0579$^{b,c}$ & \textbf{\cellcolor{lightgray}.1300}$^{c}$ & .0549$^{a,b,c}$ & .0572$^{b,c}$ & \textbf{.0985}$^{b,c}$ & .0467$^{b,c}$ & .0524$^{b,c}$ & .0670$^{b,c}$ & .0039$^{a,b}$ \\
			& p7-objective & .0386$^{a,b,c}$ & .0282$^{a,b,c}$ & \cellcolor{lightgray}.1180$^{c}$ & .0079$^{a,b,c}$ & .0405$^{a,b,c}$ & .0749$^{b,c}$ & .0335$^{a,b,c}$ & .0496$^{a,b,c}$ & .0222$^{a,b,c}$ & .0471$^{b,c}$ \\
			\midrule
			\multirow{4}{*}{\rotatebox[origin=c]{90}{F3}} & \multicolumn{11}{c}{Manual=.0497, conceptual=.0101, objective=.0023} \\
			& p6-Manual & .0794$^{b,c}$ & \textbf{.0920}$^{b,c}$ & \cellcolor{lightgray}.1440$^{c}$ & \textbf{.1196}$^{c}$ & \textbf{.0924}$^{b,c}$ & .1053$^{b,c}$ & \textbf{.0834}$^{b,c}$ & \textbf{.0957}$^{b,c}$ & \textbf{.1103}$^{b,c}$ & \textbf{.0802}$^{b,c}$ \\
			& p7-conceptual & \textbf{.1149}$^{b,c}$ & .0799$^{b,c}$ & \textbf{\cellcolor{lightgray}.1539}$^{c}$ & .0776$^{a,b,c}$ & .0780$^{b,c}$ & \textbf{.1209}$^{c}$ & .0644$^{b,c}$ & .0697$^{b,c}$ & .0881$^{b,c}$ & .0069$^{a,b}$ \\
			& p7-objective & .0475$^{a,b,c}$ & .0387$^{a,b,c}$ & \cellcolor{lightgray}.1476$^{c}$ & .0148$^{a,b,c}$ & .0582$^{b,c}$ & .1005$^{b,c}$ & .0494$^{a,b,c}$ & .0652$^{b,c}$ & .0348$^{a,b,c}$ & .0652$^{b,c}$ \\
			\bottomrule
	\end{tabular}}
\end{table*}

The reproducibility study of~\citet{Moritz2024boolean-reproduce} does not investigate Boolean query refinement, which was investigated in the original study. We investigate all tested LLMs for their ability to refine Boolean queries. Table~\ref{tab:clef-Refinement} shows the results of Boolean query refinement on the CLEF TAR collection. We confirm the original claim by~\citet{wang2023can} that Boolean query refinement by LLMs increases precision with a drop in recall. We obtain higher precision with p6-manual and p7-conceptual and higher recall with p7-conceptual. For p7-objective, the recall value is lower than the original, while precision is similar to that in the original study.

\subsection{Impact of Prompt and Output Type}

\enlargethispage{2\baselineskip}
We next investigate prompt and output types. Specifically, while most models we tested separate system prompts from user prompts, the original study used the ChatGPT interface in its early stages, where system prompts were not adjustable. We assume that at this stage, all prompts were originally fed into the user prompt. Additionally, while the~\citet{Moritz2024boolean-reproduce} study enforced JSON output for better control over response formatting, the original study used plain text. Recent literature suggests that enforcing JSON output can degrade the effectiveness of LLM-generated responses~\cite{tam2024let}.

\begin{table}
	\centering
		\caption{Results on CLEF TAR for p4 prompt variations.\vspace{-5pt}}
		\begin{tabular}{lllllll}
			\toprule
			Prompt & \multicolumn{3}{c}{GPT3.5-1} & \multicolumn{3}{c}{Mistral-S} \\ 
			& Precision & F1 & Recall & Precision & F1 & Recall \\ \cmidrule{2-7}
			Plain-text & \textbf{.1064} & \textbf{.0996} & .5108 & \textbf{.0844} & .0617 & .4831 \\ 
			JSON & .0859 & .0943 & \textbf{.5562} & .0679 & .0424 & .3914 \\ 
			System-JSON & .0898 & .0901 & .5413 & .0839 & \textbf{.0656} & \textbf{.5828} \\
			\bottomrule
		\end{tabular}

	\label{tab:p4_variations}
\end{table}

\enlargethispage{2\baselineskip}
We used plain-text prompts using the user prompt only, as system prompts and JSON output were not available for some models. We also investigated the impact of separating prompts and restricting output formats, focusing on GPT3.5-1 and Mistral-S, using p4 as the reference prompt. Table~\ref{tab:p4_variations} compares plain-text, JSON output, and System-JSON settings. The impact of JSON output varies across models. Enforcing JSON output resulted in increased recall but decreased precision and F1-score for GPT3.5-1, while Mistral-S showed a drop in all metrics. For System-JSON, we observed a varied trend: it increased both recall and precision for Mistral-S, but for GPT3.5-1, it led to a decrease in recall and an increase in precision.

Overall, none of the above results showed statistical significance. Therefore, we conclude that the use of system prompts and enforced JSON output, despite their varied effects, likely has a minimal impact on the quality of Boolean query formulation.

\subsection{Variability and Incorrect Formulation}
\begin{figure}[t!]
	\centering
	\includegraphics[width=0.8\columnwidth]{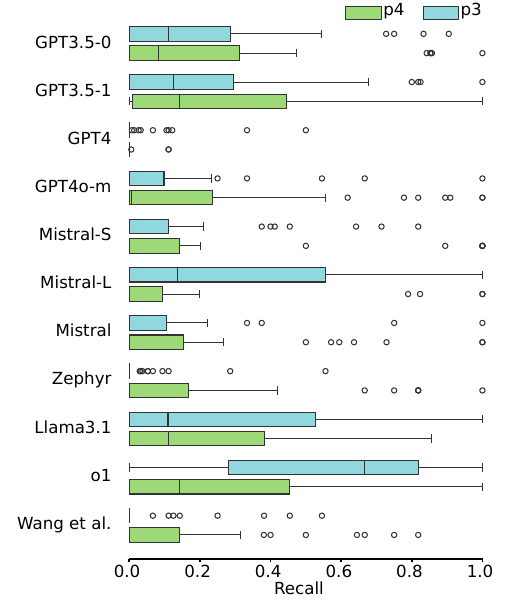}
	\caption{Recall variability of formulated Boolean queries using prompts p3 and p4 for the Seed collection.\vspace{-5pt}}
	\label{fig:variability-analysis}
\end{figure}

\begin{figure}[t!]
	\centering
	\includegraphics[width=0.8\columnwidth]{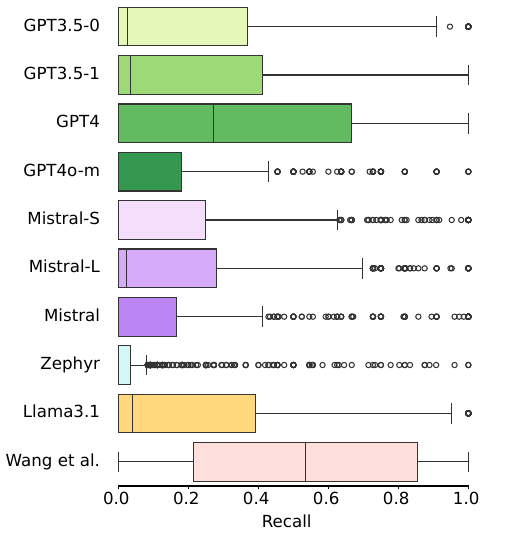}
	\caption{ Recall variability of all formulated Boolean queries using guided prompt for the Seed collection. Generation from all the seed studies is aggregated together.\vspace{-8pt}}
	\label{fig:variability-seed-analysis}
\end{figure}

To investigate the variability in Boolean query formulation effectiveness across different models, we present results for three prompts: p3, p4, and guided queries. Notably, p3 is a zero-shot prompt, p4 is a one-shot prompt, while the guided prompt uses seed studies. Figure~\ref{fig:variability-analysis} illustrates the variability of all queries generated on the Seed collection. The figure reveals high variance across models in their ability to formulate effective Boolean queries, indicating that some models are inherently unsuitable for this task.

For example, GPT-4 consistently exhibits very low recall across all queries it generates. This suggests that model selection is crucial—certain models, such as GPT-4, should be avoided when aiming for effective Boolean query formulation. In contrast, o1 models demonstrate higher effectiveness compared to all other models.

Regarding prompt selection, while most models benefit from an example to improve Boolean query formulation, certain models, such as Mistral-L and o1, actually perform worse with a one-shot prompt. Overall, the best prompt to use is model-dependent; one prompt working well on one model does not necessarily mean it would also achieve high effectiveness on other models.

Figure~\ref{fig:variability-seed-analysis} presents the variability in guided Boolean query formulation. These results indicate that, although most queries in the original study by~\citet{wang2023can} achieved higher recall and were optimised for recall, their overall effectiveness was still lower compared to the variance observed across different models, with respect to Table~\ref{tab:seed-Formulation}. This highlights the importance of seed selection for guided queries, suggesting that a model capable of generating a single high-quality Boolean query based on a seed study is more effective than one producing multiple relatively average-quality queries across all seed studies.

\begin{figure}[t]
	\centering
	\includegraphics[width=0.7\columnwidth]{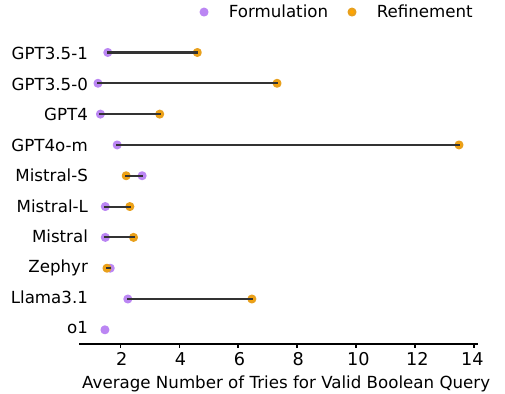}
	\caption{Average number of tries per model to generate the first valid Boolean query. o1 was not used for query refinement due to high cost.}
	\label{fig:falling}
\end{figure}

Figure~\ref{fig:falling} presents the average number of attempts required for each model to generate the first valid Boolean query. The results indicate that GPT models generally needed fewer attempts, with the notable exception of GPT4o-m, which required the highest number of retries. Additionally, Mistral-S exhibited the highest number of retries for query formulation prompts.

Boolean query refinement required more attempts to produce a valid query, likely due to the added complexity of both modifying the existing query and ensuring its correctness. This gap is particularly large for smaller GPT-based models, such as GPT-3.5 and GPT-4o-mini, which struggle more with refinement compared to formulation. However, for Mistral-based models and most open-source models (except Llama3.1), the number of retries for refinement does not appear to be substantially higher than for formulation. Notably, models such as Mistral-S and Zephyr required fewer retries on average for refinement than for initial query formulation.

\begin{table*}[ht]
	\centering
	\tiny
	\setlength\tabcolsep{0pt}
	\renewcommand{\arraystretch}{1.2}
	\caption{Comparison of different models generating Boolean queries for topic 22 from the Seed collection, using \textit{p4}.}
		\label{tbl:case_study_models}
	\begin{tabular*}{\linewidth}{@{\extracolsep{\fill}}lp{320pt}llll}
		\toprule
		Model & Boolean Query & P & F1 & F3 & R \\
		\midrule
		Manual & ("rabies vaccines"[Mesh] OR ((vaccination[Mesh]) AND (rabies[Mesh] OR "rabies virus"[Mesh])) OR ((Rabies[tiab]) AND (vaccine[tiab] OR Vaccines[tiab] OR Vaccination[tiab] OR Vaccinations[tiab] OR Immunise[tiab] OR Immunize[tiab] OR Immunisation[tiab] OR Immunization[tiab]))) AND ("immunization, secondary"[Mesh] OR "antibodies, viral"[Mesh] OR "time factors"[Mesh] OR secondary[tiab] OR booster[tiab] OR boost[tiab] OR boosters[tiab] OR boostability[tiab] OR titer[tiab] OR gmt[tiab] OR ((sero[tiab]) AND (conversion[tiab] OR protection[tiab])) OR seroconversion[tiab] OR seroprotection[tiab]) AND ("administration and dosage"[sh] OR Administer[tiab] OR Administered[tiab] OR Administration[tiab] OR Intramuscular[tiab] OR Intradermal[tiab] OR “pre-exposure”[tiab] OR “pre exposure”[tiab]) NOT (animals[Mesh] NOT humans[Mesh])  & 0.113 & 0.201 & 0.332 & 0.947 \\ \midrule
		GPT3.5-1 & ("rabies vaccines"[MeSH Terms] OR "rabies vaccines"[All Fields] OR "rabies vaccination"[All Fields] OR "rabies vaccine"[All Fields] OR "rabies immunization"[All Fields] OR "immunization"[MeSH Terms] OR "immunization"[All Fields]) AND ("immunogenicity"[MeSH Terms] OR "immunogenicity"[All Fields]) AND ("dose-response relationship, immunologic"[MeSH Terms] OR "dose-response relationship, immunologic"[All Fields] OR "dose-response"[All Fields] OR "dose response"[All Fields] OR "dose-effect"[All Fields] OR "dose effect"[All Fields]) & 0.003 & 0.006 & 0.010 & 0.026 \\ \midrule
		GPT3.5-0 & ("rabies vaccines"[MeSH Terms] OR "rabies vaccines"[All Fields] OR "rabies vaccine"[All Fields] OR rabies[All Fields]) AND ("immunogenicity"[MeSH Terms] OR "immunogenicity"[All Fields] OR "immunogenicity"[MeSH Terms] OR "immunogenicity"[All Fields] OR "immune response"[MeSH Terms] OR "immune response"[All Fields]) AND ("dose-response relationship, drug"[MeSH Terms] OR "dose-response relationship, drug"[All Fields] OR "dose-response"[All Fields] OR "dose response"[All Fields] OR "dose dependent"[All Fields]) & 0.065 & 0.037 & 0.031 & 0.026 \\ \midrule
		GPT4 & (rabies[mesh] OR rabies[All Fields]) AND (vaccin*[mesh] OR vaccin*[All Fields] OR immuniz*[mesh] OR immuniz*[All Fields] OR immunisation[All Fields] OR immunization[All Fields]) AND (immunogenic*[mesh] OR immunogenic*[All Fields]) AND ("pre-exposure prophylaxis"[mesh] OR pre-exposure[All Fields] OR "post-exposure prophylaxis"[mesh] OR post-exposure[All Fields]) AND (systematic[sb] AND (Review[ptyp] OR meta-analysis[ptyp] OR meta-analysis as topic[mesh] OR meta-analysis[All Fields])) & 0.000 & 0.000 & 0.000 & 0.000 \\ \midrule
		GPT4o-m & immunogenicity[All Fields] AND (rabies[mesh:noexp] OR rabies[All Fields] OR rabies vaccine[All Fields] OR rabies vaccinations[All Fields] OR rabies immunization[All Fields] OR rabies prophylaxis[All Fields]) AND (pre-exposure[All Fields] OR post-exposure[All Fields] OR dose-response[All Fields] OR (vaccination[All Fields] AND response[All Fields])) & 0.075 & 0.134 & 0.221 & 0.618 \\ \midrule
		Mistral-S & (Rabies[Mesh] OR Rabies Vaccines[Mesh] OR Rabies Immunization[Mesh] OR "Rabies vaccine"[All Fields] OR "pre-exposure prophylaxis"[All Fields] OR "post-exposure prophylaxis"[All Fields] OR "Rabies PEP"[All Fields] OR "Rabies POEP"[All Fields] OR rabies prophylaxis[All Fields] OR immunogenicity[All Fields] OR antibody response[All Fields] OR seroconversion[All Fields] OR vaccine efficacy[All Fields] OR "dose-response"[All Fields]) AND ("Systematic Review"[PB] OR "Meta-Analysis"[Mesh] OR "Review Literature as Topic"[Mesh] OR "Review Literature"[All Fields] OR "Systematic Literature Review"[All Fields]) & 0.000 & 0.000 & 0.000 & 0.000 \\ \midrule
		Mistral-L & ("Rabies Vaccination"[Mesh] OR "Rabies Vaccines"[Mesh] OR "Rabies Vaccines, Inactivated"[Mesh] OR "Rabies Vaccines, Live"[Mesh] OR "Rabies Immunization"[Mesh] OR "Rabies Immunoglobulin, Human"[Mesh] OR "Rabies Immunoglobulins"[Mesh] OR "Rabies Virus Neutralizing Antibodies"[Mesh] OR "Rabies Vaccines, Human Diploid Cell"[Mesh] OR "Rabies Vaccines, Vertebrate Cell Culture"[Mesh] OR "Rabies"[All Fields] OR "rabies vaccines"[All Fields] OR "rabies vaccine"[All Fields] OR "rabies vaccination"[All Fields] OR "rabies immunization"[All Fields] OR "rabies prophylaxis"[All Fields] OR "post-exposure prophylaxis"[All Fields] OR "post-exposure prophylaxis, rabies"[All Fields] OR "pre-exposure prophylaxis"[All Fields] OR "pre-exposure prophylaxis, rabies"[All Fields] OR "rabies immunoglobulin"[All Fields] OR "rabies immunoglobulins"[All Fields] OR "rabies ig"[All Fields] OR "rabies igg"[All Fields] OR "rabies antibodies"[All Fields] OR "rabies titer"[All Fields] OR "rabies serology"[All Fields] OR "rabies seronegative"[All Fields] OR "rabies seropositive"[All Fields]) AND (Immunogenicity[All Fields] OR "immune response"[All Fields] OR "antibody response"[All Fields] OR "antibody titer"[All Fields] OR "antibody production"[All Fields] OR "immune protection"[All Fields] OR "immunization response"[All Fields] OR "vaccine response"[All Fields]) AND (Systematic[All Fields] OR "Meta-Analysis"[Publication Type] OR "meta analysis"[All Fields] OR "meta-analyses"[All Fields] OR "review"[Publication Type] OR "review, systematic"[Publication Type] OR "systematic reviews"[All Fields]) & 0.002 & 0.004 & 0.006 & 0.013 \\ \midrule
		Mistral & ("Rabies Vaccines/" OR "rabies vaccination"[All Fields] OR "rabies immunization"[All Fields]) AND (Immunogenicity[Mesh] OR "antibody response"[All Fields] OR "immune response"[All Fields] OR "antigen-antibody complexes"[Mesh] OR "humoral immunity"[Mesh]) AND ("pre-exposure prophylaxis"[Mesh] OR "post-exposure prophylaxis"[Mesh] OR "immunization, postexposure"[Mesh]) & 0.207 & 0.114 & 0.093 & 0.079 \\ \midrule
		Zephyr & ("rabies vaccine"[Mesh] OR ("rabies"[All Fields] AND "vaccine"[All Fields])) AND (("immunogenicity"[Mesh] OR "immunogenicity"[All Fields]) OR ("antibody response"[Mesh] OR "antibody response"[All Fields])) AND (("pre-exposure"[All Fields] OR "pre exposure"[All Fields]) OR ("post-exposure"[All Fields] OR "post exposure"[All Fields])) AND (("systematic review"[All Fields] OR "meta-analysis"[All Fields]) ) & 0.000 & 0.000 & 0.000 & 0.000 \\ \midrule
		Llama3.1 & (rabies vaccine*[All Fields] OR post exposure prophylaxis[Mesh:noexp] OR pre exposure prophylaxis[Mesh:noexp] OR rabies immunoglobulin[Mesh:noexp] OR (rabies[All Fields] AND immunogen*[All Fields]) OR (rabies[All Fields] AND vaccin*[All Fields]) OR ("rabies in situ"[All Fields] OR post vaccine react*[All Fields] OR vaccine reaction*[All Fields])) AND (immunogenicity[Mesh:noexp] OR immunogenecity[All Fields] OR immune respons*[All Fields] OR "dose response"[All Fields] OR "dose response relationship"[All Fields] OR "dose response curve"[All Fields] OR (immunogen*[All Fields] AND response[All Fields] AND dose[All Fields])) & 0.046 & 0.083 & 0.138 & 0.421 \\ \midrule
		o1 & (rabies[MeSH:noexp] OR "Rabies Virus"[MeSH:noexp] OR "Rabies Vaccines"[MeSH:noexp] OR rabies[tiab] OR "rabies virus"[tiab] OR "rabies infection"[tiab] OR "rabies vaccine"[tiab]) AND (("Immunization, Pre-Exposure"[MeSH:noexp] OR ((pre-exposure*[tiab] OR preexposure*[tiab]) AND (prophylaxis[tiab] OR vaccinat*[tiab] OR immuni*[tiab]))) OR ("Immunization, Post-Exposure"[MeSH:noexp] OR ((post-exposure*[tiab] OR postexposure*[tiab]) AND (prophylaxis[tiab] OR vaccinat*[tiab] OR immuni*[tiab])) OR PEP[tiab])) AND ("Immunogenicity, Vaccine"[MeSH:noexp] OR immunogenic*[tiab] OR (immune[tiab] AND (response*[tiab] OR respons*[tiab])) OR "antibody formation"[MeSH:noexp] OR seroconversion*[tiab] OR (antibody[tiab] AND (titer*[tiab] OR titre*[tiab] OR level*[tiab] OR concentration*[tiab])) OR ((neutralizing[tiab] OR neutralising[tiab]) AND (antibody[tiab] OR antibodies[tiab]))) AND humans[MeSH Terms] & 0.141 & 0.241 & 0.372 & 0.816 \\ \midrule
	\end{tabular*}
	\footnotesize{Topic title: \textit{Immunogenicity after pre- and post-exposure rabies vaccination: A systematic review and dose-response meta-analysis}}
\end{table*}

\begin{table*}[ht]
	\centering
		\tiny
	\setlength\tabcolsep{0pt}
	\renewcommand{\arraystretch}{1.2}
	\caption{Comparison of different prompts used to generate Boolean queries for topic 22 from the Seed collection, using o1.}
		\label{tbl:case_study_prompts}
	\begin{tabular*}{\linewidth}{@{\extracolsep{\fill}}l p{320pt} llll}
		\toprule
		Prompt & Boolean Query & P & F1 & F3 & R \\
		\midrule
		Manual & ("rabies vaccines"[Mesh] OR ((vaccination[Mesh]) AND (rabies[Mesh] OR "rabies virus"[Mesh])) OR ((Rabies[tiab]) AND (vaccine[tiab] OR Vaccines[tiab] OR Vaccination[tiab] OR Vaccinations[tiab] OR Immunise[tiab] OR Immunize[tiab] OR Immunisation[tiab] OR Immunization[tiab]))) AND ("immunization, secondary"[Mesh] OR "antibodies, viral"[Mesh] OR "time factors"[Mesh] OR secondary[tiab] OR booster[tiab] OR boost[tiab] OR boosters[tiab] OR boostability[tiab] OR titer[tiab] OR gmt[tiab] OR ((sero[tiab]) AND (conversion[tiab] OR protection[tiab])) OR seroconversion[tiab] OR seroprotection[tiab]) AND ("administration and dosage"[sh] OR Administer[tiab] OR Administered[tiab] OR Administration[tiab] OR Intramuscular[tiab] OR Intradermal[tiab] OR “pre-exposure”[tiab] OR “pre exposure”[tiab]) NOT (animals[Mesh] NOT humans[Mesh])  & 0.113 & 0.201 & 0.332 & 0.947 \\ \midrule
		p1 & ("rabies"[MeSH Terms] OR rabies[Title/Abstract] OR "rabies virus"[Title/Abstract]) AND ("rabies vaccines"[MeSH Terms] OR "rabies vaccine"[Title/Abstract] OR "rabies vaccination"[Title/Abstract] OR vaccine*[Title/Abstract]) AND (immunogenic*[Title/Abstract] OR "immune response"[Title/Abstract] OR "immunological response"[Title/Abstract]) AND (("pre-exposure prophylaxis"[MeSH Terms] OR "pre-exposure prophylaxis"[Title/Abstract] OR preexposure[Title/Abstract]) OR ("post-exposure prophylaxis"[MeSH Terms] OR "post-exposure prophylaxis"[Title/Abstract] OR postexposure[Title/Abstract])) & 0.222 & 0.326 & 0.427 & 0.618 \\ \midrule
		p2 & ("Rabies Vaccines"[Mesh] OR "rabies vaccine*"[tiab] OR "rabies vaccination*"[tiab]) AND ("pre-exposure prophylaxis"[tiab] OR "preexposure prophylaxis"[tiab] OR "pre exposure prophylaxis"[tiab] OR "post-exposure prophylaxis"[tiab] OR "postexposure prophylaxis"[tiab] OR "post exposure prophylaxis"[tiab]) AND ("immunogenicit*"[tiab] OR "immune respons*"[tiab] OR "immunologic respons*"[tiab] OR "serologic respons*"[tiab] OR "antibody titer*"[tiab] OR "antibody titre*"[tiab] OR seroconversion[tiab] OR "virus neutralizing antibody*"[tiab] OR "virus neutralising antibody*"[tiab]) & 0.213 & 0.286 & 0.345 & 0.434 \\ \midrule
		p3 & (("rabies"[MeSH Terms] OR "rabies virus"[MeSH Terms] OR rabies[tiab] OR "rabies virus"[tiab]) AND ("rabies vaccines"[MeSH Terms] OR "rabies immunoglobulin"[MeSH Terms] OR "vaccination"[MeSH Terms] OR "pre-exposure prophylaxis"[MeSH Terms] OR "post-exposure prophylaxis"[MeSH Terms] OR "rabies vaccine*"[tiab] OR "rabies immunoglobulin"[tiab] OR vaccin*[tiab] OR immuniz*[tiab] OR immunis*[tiab] OR "pre-exposure prophylaxis"[tiab] OR "post-exposure prophylaxis"[tiab] OR PEP[tiab] OR PrEP[tiab] OR RIG[tiab] OR HRIG[tiab]) AND (immunogenicity[tiab] OR "immune response*"[tiab] OR "antibody titer*"[tiab] OR "antibody titre*"[tiab] OR antibodies[tiab] OR seroconversion[tiab] OR seroconvert*[tiab] OR "humoral immunity"[tiab] OR "immune memory"[tiab] OR "neutralizing antibody*"[tiab] OR "neutralising antibody*"[tiab] OR "virus neutralizing antibody*"[tiab])) NOT (animals[MeSH Terms] NOT humans[MeSH Terms]) & 0.073 & 0.136 & 0.239 & 0.987 \\ \midrule
		p4 & (rabies[MeSH:noexp] OR "Rabies Virus"[MeSH:noexp] OR "Rabies Vaccines"[MeSH:noexp] OR rabies[tiab] OR "rabies virus"[tiab] OR "rabies infection"[tiab] OR "rabies vaccine"[tiab]) AND (("Immunization, Pre-Exposure"[MeSH:noexp] OR ((pre-exposure*[tiab] OR preexposure*[tiab]) AND (prophylaxis[tiab] OR vaccinat*[tiab] OR immuni*[tiab]))) OR ("Immunization, Post-Exposure"[MeSH:noexp] OR ((post-exposure*[tiab] OR postexposure*[tiab]) AND (prophylaxis[tiab] OR vaccinat*[tiab] OR immuni*[tiab])) OR PEP[tiab])) AND ("Immunogenicity, Vaccine"[MeSH:noexp] OR immunogenic*[tiab] OR (immune[tiab] AND (response*[tiab] OR respons*[tiab])) OR "antibody formation"[MeSH:noexp] OR seroconversion*[tiab] OR (antibody[tiab] AND (titer*[tiab] OR titre*[tiab] OR level*[tiab] OR concentration*[tiab])) OR ((neutralizing[tiab] OR neutralising[tiab]) AND (antibody[tiab] OR antibodies[tiab]))) AND humans[MeSH Terms] & 0.141 & 0.241 & 0.372 & 0.816 \\ \midrule
		p5 & ("Rabies Vaccines"[Mesh:noexp] OR (rabies[All Fields] AND vaccin*[All Fields]) OR "rabies vaccine"[All Fields] OR "rabies immunoglobulin"[All Fields] OR rabipur[All Fields] OR verorab[All Fields] OR imovax[All Fields] OR "pre-exposure prophylaxis"[All Fields] OR "post-exposure prophylaxis"[All Fields] OR "rabies prophylaxis"[All Fields]) AND ("Immunogenicity, Vaccine"[Mesh:noexp] OR immunogenic*[All Fields] OR immunolog*[All Fields] OR (antibod*[All Fields] AND (respons*[All Fields] OR titer*[All Fields] OR titre*[All Fields])) OR seroconversion[All Fields] OR "immune response"[All Fields] OR (virus[All Fields] AND neutraliz*[All Fields]) OR immunity[Mesh]) & 0.017 & 0.034 & 0.066 & 1.000 \\ \midrule
	\end{tabular*}
	\footnotesize{Topic title: \textit{Immunogenicity after pre- and post-exposure rabies vaccination: A systematic review and dose-response meta-analysis}}
\end{table*}

\section{Case Study}

We present a case study using one exemplar topic to compare (1) the Boolean queries formulated by different LLMs when the same prompt is used (p4, Table~\ref{tbl:case_study_models}), and (2) the Boolean queries formulated by different prompts when the same LLM is used (o1, Table~\ref{tbl:case_study_prompts}). We use topic 22 from the Seed collection~\footnote{This topic is chosen due to the varying effectiveness observed when using different LLMs and prompts.}.

\subsection{Comparison by LLMs}

Boolean queries generated by different LLMs vary notably in structure, term coverage, and retrieval effectiveness. The manually formulated query remains the most effective, combining MeSH terms and diverse free-text synonyms to maximise recall. For example:

\begin{grayquote}
	\raggedright
	\ttfamily \small
	(“rabies vaccines”[Mesh] OR “rabies virus”[Mesh] OR Rabies[tiab] AND (vaccine[tiab] OR Immunization[tiab])) AND (…)
\end{grayquote}

\noindent
Among the automated methods, \textbf{o1} generated the most effective queries. Its outputs captured the core concepts with precise Boolean structure, selectively applied filtering, and maintained broad term coverage without unnecessary restrictions:

\begin{grayquote}
	\raggedright
	\ttfamily \small
	(rabies[MeSH:noexp] OR “Rabies Virus”[MeSH:noexp] OR rabies[tiab]) AND (immune response[tiab] OR immunogenicity[MeSH]) AND humans[MeSH Terms]
\end{grayquote}

\noindent
\textbf{GPT-4o-m} and \textbf{Llama 3.1} produced moderately effective queries. Both models achieved reasonable term coverage and applied some structuring, but their outputs remained either overly simplified (GPT-4o-m) or loosely connected (Llama 3.1), leading to weaker retrieval effectiveness compared to o1.

In contrast, \textbf{GPT-3.5} generated queries that were overly simplistic, with insufficient expansion and minimal domain adaptation, resulting in poor retrieval performance.

Several models, including \textbf{GPT-4}, \textbf{Mistral}, and \textbf{Zephyr}, severely underperformed by incorrectly embedding rigid constraints, such as requiring retrieved documents to explicitly mention “systematic review” or “meta-analysis.” Such constraints are fundamentally misaligned with the retrieval objective: they drastically narrow the retrieved set and often result in zero relevant documents being retrieved. This reflects a critical failure in adapting query generation to retrieval needs.

\textit{In summary, effective automated Boolean query generation requires comprehensive expansion of relevant concepts, precise logical structuring, and avoidance of overly restrictive filters. Models like o1 demonstrate that striking the right balance between recall and precision is essential for generating high-quality queries automatically.}

\subsection{Comparison by Prompts}

Boolean queries also vary significantly based on the prompt used, even when generated by the same model. Different prompts emphasize different aspects of query construction, influencing recall and precision. For instance, prompts like p1 and p2 are concise and structured, prioritizing Boolean logic but limiting term expansion. The query generated from p1 is:

\begin{grayquote}
\raggedright
\ttfamily \small
("rabies"[MeSH Terms] OR rabies[Title/Abstract] OR "rabies virus"[Title/Abstract]) AND ("rabies vaccines"[MeSH Terms] OR "rabies vaccine"[Title/Abstract]) AND (immunogenic*[Title/Abstract])
\end{grayquote}

\noindent
In contrast, p3 and p4 encourage broader term expansion, leading to more inclusive queries. The query from p4 includes:
\begin{grayquote}
\raggedright
\ttfamily \small
(rabies[MeSH:noexp] OR "Rabies Virus"[MeSH:noexp]) AND ("Immunization, Pre-Exposure"[MeSH:noexp] OR pre-exposure*[tiab]) AND (immunogenic*[tiab] OR "antibody formation"[MeSH:noexp])
\end{grayquote}
\noindent
\enlargethispage{2\baselineskip}
p5, which uses PICO (Patient, Intervention, Comparison, Outcome), results in the most structured queries, balancing precision and recall. Overall, prompts designed for expert systematic review specialists (p3, p4, p5) tend to generate more comprehensive queries, whereas direct and simple prompts (p1, p2) lead to narrower searches. The effectiveness of a prompt depends on the balance between recall and specificity needed for a systematic review.

\section{Discussion and Limitations}

We highlighted three key aspects of the study by~\citet{Moritz2024boolean-reproduce} that differed from the original study of~\citet{wang2023can}. Across these three aspects, we found that:
\begin{enumerate*}
	\item query validation plays the biggest role in the difference in query formulation effectiveness observed by~\citet{Moritz2024boolean-reproduce};
	\item seed selection plays the biggest role in the difference in guided query formulation effectiveness observed by~\citet{Moritz2024boolean-reproduce}; and
	\item the output format plays an insignificant role in the difference in effectiveness of all methods reproduced by~\citet{Moritz2024boolean-reproduce}.
\end{enumerate*}
In properly validating generated queries and choosing appropriate seed studies, we were able to replicate the results of~\citet{wang2023can}.

One unexpected result that we found was that the effectiveness of many methods was higher (often statistically significantly) than the original results by~\citet{wang2023can}. It is unclear whether these results were due to improvements in model architecture or training data. Given the fact that the test collections used in this paper and the systematic reviews underlying them are openly available on the web, this increase in effectiveness could be due to them ending up in the training data for many of these models. For future work, we will investigate prospectively evaluating the methods from this study using the approach outlined by~\citet{kusa2024leveraging}.

	\section{Conclusion}

We systematically investigated the effectiveness of LLMs for Boolean query formulation and refinement in systematic review literature search. By revisiting~\citet{wang2023can} and addressing key aspects overlooked in~\citet{Moritz2024boolean-reproduce}, we provided a more comprehensive evaluation of LLM-based Boolean query generation.

Our findings show that query formulation effectiveness varies significantly across models and prompt designs. While newer models, such as o1, generally perform better, results remain inconsistent. Boolean query refinement improves precision but often reduces recall, especially for smaller models, which require more attempts to generate valid queries. We also find that output formatting (plain-text vs. JSON) has minimal impact, whereas model selection and prompt design play a more critical role.

Additionally, we find that guided formulation benefits from well-chosen seed studies, supporting~\citet{wang2023can} but contradicting~\citet{Moritz2024boolean-reproduce}. As seed selection is post-hoc, we explored combining queries from multiple seeds, which improved recall at the cost of precision, offering a potential method for generating high-recall Boolean queries. Additionally, some models, such as GPT-4o-m, frequently fail to generate valid queries, requiring multiple retries.

Overall, LLMs show promise for Boolean query generation, but their effectiveness depends on model selection and prompt strategies. Future research should refine prompt engineering, explore domain-specific optimizations, and develop hybrid approaches integrating statistical and generative methods. The reproduction code and generated queries from this study are available at\linebreak \url{https://github.com/ielab/Boolean_Generation_Reproduce}.

\paragraph{Acknowledgement}
Shuai Wang is supported by a UQ Earmarked PhD Scholarship. This research is funded by the Australian Research Council Discovery Project DP210104043.

	% Uncomment if needed
	% \section*{Acknowledgments}
	% Shuai Wang is supported by a UQ Earmarked PhD Scholarship. This research is funded by a Universities Australia--DAAD Joint Research Co-Operation Scheme and the Australian Research Council Discovery Project DP210104043.
	
	\bibliographystyle{plainnat}
	\interlinepenalty=10000
	\bibliography{bibliography}
	
\end{document}